\documentclass{ws-ijmpe}

\usepackage{graphicx}
\begin{document}

\markboth{Authors' Names}{Summary talk}
\catchline{}{}{}{}{}

\title{HEAVY ION DYNAMICS AND NEUTRON STARS}
\author{MASSIMO DI TORO $^*$}%
\address{Laboratori Nazionali del Sud INFN, I-95123 
Catania, Italy,\\
and Physics-Astronomy Dept., University of Catania\\%
$^*$ E-mail:\_ditoro@lns.infn.it}

\maketitle
\begin{history}%
\received{October 07}%
\revised{October 07}%
\end{history}

\begin{abstract}
Some considerations are reported, freely inspired from the presentations
and discussions during the Beijing Normal University Meeting on the above
 Subject,
held in July 2007. Of course this cannot be a complete summary but
 just a collection
of personal thougths aroused during the workshop.
\end{abstract}

\section{Introduction}

It is a real pleasure to write a few summary notes on this very
 stimulating 
International Workshop held in the nice atmosphere of the Normal University
 of Beijing. First of all I have to stress the interesting project that 
was behind the organization of such meeting: try to put together experts 
working in apparently distant fields of the nuclear research. The Heavy Ion 
Dynamics studies (in terrestrial laboratories) address typical non-equilibrium
properties of nuclear systems. The Compact Stars are objects 
(not directly probed) which show typical equilibrium properties of a 
nuclear matter very far away from normal conditions. The link is in fact
extremely clear and based on two fundamental topics that have been 
extensively discussed during the workshop:
\begin{itemize} 
\item{1. The structure of the effective nuclear interactions in extreme 
conditions for various degrees of freedom, density, isospin, spin, 
strangeness....}
\item{2. The phases of nuclear matter, with particular attention to the high
temperature and/or high density phase transitions, chiral vs. deconfinement.}
\end{itemize}

Before passing to a more detailed analysis of the main points discussed at 
the meeting I would like to make a general comment on the fact that this 
clear connection between Heavy ion Collisions ($HIC$) and 
Neutron Stars ($NS$) is not much present in the activity of the 
Radioactive Ion Beam ($RIB$) Community: the main focus is on nuclear
structure problems, studied with "soft" peripheral collisions. More 
dissipative reactions with exotic beams, testing the Equation of State ($EOS$)
of asymmetric matter, seem not to be of much interest. We are at the level of
a dogmatic "Doctrina Una Veritas Una" ("One Doctrine One Truth"), 
not really a scientific approach.

\section{Neutron Stars: A Many-EOS Problem}

The relationship between equilibrium properties of $NS$
(mass/radius, central densities, cooling...) and the $EOS$ of a "Nuclear"
matter in various phases (hadronic, mixed, partonic) has emerged very 
clearly. Unfortunately we have not yet reached the knowledge of an $unique$ 
$EOS$ which can consistently describe the hadron-quark transition.
Very likely such transition is taking place in the interior of some compact 
stars. In order to make predictions we must refine our knowledge of hadronic 
and partonic $EOS$'s and suggest some sensitive observables.

\subsection*{Hadronic EOS}

The inclusion of 3-body forces is essential in order to get the correct 
saturation point for symmetric $NM$. $U.Lombardo$ has shown that in a 
non-relativistic Brueckner-Hartree-Fock ($BHF$) approach this contribution 
leads
to a more repulsive matter at high baryon and isospin density
\cite{zhuoprc69}. 
 The latter 
point, on a more repulsive isospin potentials (symmetry energy) at high
density, appears very important for various aspects: 
\begin{itemize} 
\item{ We can reconcile a "stiff" $EOS$ for asymmetric matter with the
relatively "soft" behavior required in isospin symmetric cases from
relativistic heavy ion collisions, in order to account for nucleon/meson
production and flows \cite{daniel_sci02}. }
\item{ In a pure hadronic picture a large symmetry repulsion will 
allow large values for the maximum mass that can be reached by a $NS$, up 
to more 
than 2 solar masses ($M_\odot$) . This appears consistent with some recent 
observations \cite{page06}.}
\item{ A large symmetry repulsion will favor an earlier transition to a mixed 
hadron-quark phase at lower baryon densities in the very neutron rich matter
of $NS$s.}
\end{itemize}

During the discussions the importance of Exchange Terms (the Fock Terms) 
in the 
evaluation of spin-isospin contribution to the nuclear $EOS$ has been stressed
many times. I have a comment on this point for the young students, a kind 
of paradox.
{\em In all Skyrme effective interactions the isospin dependence is
COMPLETELY coming from Exchange Terms}. No new isovector parameters are needed!

A common problem for all these repulsive hadronic $EOS$s in neutron rich matter
is the fact that the limit of a direct ($p+e^- \rightarrow n+\nu_e$) $URCA$
fast neutrino cooling
process (proton fraction $Z/A\simeq 0.11$) can be reached already at 
relatively 
low densities, around $2-3\rho_0$. The point is that we can expect 
various $NM$ 
phases at high baryon density in the inner core and this will influence 
the maximum 
masses as well as the cooling of a $NS$. Many talks have been devoted to this
topic. We must note how the various "paths" of the matter are depending on the 
scarcely known behavior 
of the symmetry energy at high density.

In this respect a nice result has been presented by $L. Bo$ within a
 Relativistic 
Mean Field
 ($RMF$) approach. In an hadron+lepton scheme the equilibrium Mass-Radius 
relation 
is largely
influenced by the effective fields in the isovector channels, even at the 
level 
of a
different density dependence of the couplings. The maximum masses appear not 
much affected,
always around $2M_\odot $, but for lower masses we can have very different 
corresponding
$NS$ radii \cite{liubo07} . This also shows the importance of more data on the 
mass/radius relation for a large 
spectrum of $NS$ masses.

The onset of Hyperon Matter has been discussed in several talks 
($U.Lombardo$, $L.Dang$,
 $T.Maruyama$, $T.Takatsuka$). It will happen when the nucleon chemical 
potentials will
reach the value of the hyperon ones. We remember that hyperons do not have 
the Fermi
kinetic contribution and so such transition is favored. The symmetry energy 
is also important,
e.g. to form lambdas from neutrons the conditon $\mu_n=\mu_\Lambda$ is better 
satisfied if 
the neutrons see a more repulsive symmetry potential. The problem is that when 
hyperons are 
formed the matter becomes much softer since we have new degress of freedom and 
an attractive
hyperon interaction \cite{burgio00,vidana06}. The result is that the expected 
maximum masses are within
two limits, a lower one around $1.3M_\odot$,
 in order to reach high densities in the inner core, and
an upper one below $1.5M_\odot$.

In fact the $YY$ interaction is not well known, in particular in the nuclear 
medium.
$T.Takatsuka$ was mentioning the $NAGARA$ event, \cite{taka01} pointing to a 
smaller attraction
in the $\Lambda \Lambda$ system, as a nice indication of some more 
repulsive terms.

Kaon ($K^-$) condensation is also expected when 
$\mu_n - \mu_p = \mu_e = \mu_K$ ($T.Maruyama$).
Again the symmetry energy is important since we have also  
$\mu_n-\mu_p=4E_{sym}(\rho)$
\cite{baranPR}. Effects on neutrino cooling are predicted via direct URCA
processes of baryonic quasiparticles \cite{nscool07}. 
A fancy $K-Pasta$ structure
has been suggested by $T.Maruyama$ due to coulomb effects. Observables?

\subsection*{Quark EOS and Hybrid Stars}

Many talks were addressing the possibility of a transition to a 
deconfined quark
matter in the inner core 
($U.Lombardo$, $T.Maruyama$, $X.Huang$, $C.Y.Gao$,$M.Di~Toro$). 
Here we directly see the uncertainty due to our poor knowledge of a 
partonic $EOS$.
The consequences are already at the level of the corresponding $NS$ 
maximum masses
for the relative $Hybrid-Stars$. 

Classes of quark $EOS$s that 
privilege the confinement,
like the MIT Bag Models, can sustain even masses of $\sim 2M_\odot$ 
while the
$NJL$-like models, that have the correct chiral properties, give 
instability conditions
for masses above $\sim 1.5M_\odot $. The inclusion of pairing correlations 
in the quark phase
are also influencing the equilibrium properties, as nicely shown 
by $C.Y.Gao$.

The impression is that we are approaching an important convergence 
on some more fundamental
effective field models, like the $NJL$, just introducing new terms that 
allow the confinement-
deconfinement transition. In this way the compact star structures will 
supply a very interesting arena 
for testing effective-QCD theories. A good review of the work at 
the Peking University
in the same direction has been presented by $Y.X.Liu$, including some 
first results with
the Polyakov-Loop implemented $NJL$ model, $P-NJL$.
Again we see an effect of the symmetry repulsion, that can lead to 
an "earlier" (lower density)
transition to a mixed hadron-quark phase in the core \cite{ditoro06}.

From the point of view of the expected neutron cooling $X.Huang$ has 
nicely shown how a direct
URCA process ($u+e\rightarrow d+\nu_e$) can develop in the quark 
phase

In conclusion a picture is emerging of roughly two main classes of neutron 
stars, one of larger
maximum mass (around $2M_\odot$) and faster cooling, nucleonic stars or with 
a repulsive 
new phase in the inner 
core, the second with smaller maximum masses (below $1.5M_\odot $) 
and slow cooling, mostly due
to modified URCA processes or nucleon-neutrino bremsstrahalung. A
 corresponding different
evolutionary scenario has been recently proposed by P.Haensel 
\cite{haensel07}. 

\subsubsection*{$NS$ Crust}

Effects of pairing on the low density surface region of a $NS$ or of a very 
heavy nuclear 
system have been addressed by $J.G.Cao$ using a $RMF$ approach coupled to 
a Gogny 
(non relativistic) treatment of the pairing interaction. The results for 
the neutron gas 
distribution are intersting but the use of the $RMF$ parametrizations at such 
low densities
($\rho\sim 10^{-3}\rho_0$) has to be justified. We remind that the free 
couplings at zero 
densities are
not recovered in the model.

$T.Maruyama$ has also presented a very interesting study of the $Pasta$ 
nuclear structures 
in the $NS$ crust expected to 
minimize the total energy
in a competition between surface and coulomb effects, again in a $RMF$ 
nuclear scheme.
We see a nice transition from droplet to rod, slab, tube and bubble non 
uniform structures.
I would like to remind that the true Italian Pasta has presently 137 (
the fine structure
constant?) different topological
structures, better minima could be certainly obtained!

\section{Constraining the Symmetry Energy with Heavy Ion Collisions}

In a meeting on Heavy Ion Collisions and Neutron Stars the Symmetry 
Energy represents
a key issue. In fact the analysis of $HIC$ data is the only tool to 
extract 
Symmetry Energy information in region far away from saturation, in 
particular 
at the low and high densities of interest for Neutron Stars, as seen before. 
However the access to the density dependence of the symmetry energy in 
terrestrial laboratory
is limited by three main reasons: i) the reduced isospin asymmetry of the 
available 
interacting nuclear systems that decreases the dynamical effect of the 
symmetry term;
ii) the limits on the density regions that can be probed, up to $2-3\rho_0$;
 iii)the presence 
of sequential decays from the primary reaction products that can severely 
wash up
the isospin signal.

For all these reasons we need a cross check of many isospin sensitive 
observables
before drawing any firm conclusion. The situation appears rather 
analogous to the 
case of the Liquid-Gas Phase Transition in $NM$. We have required more 
than ten indipendent
observables before confirming the effect.
Along this line many new ideas have been presented at the conference.

\subsection*{Probing the Symmetry Energy at Low Density}

For the study of the symmetry energy at sub-saturation densities several 
observables have been
suggested via dissipative heavy ion collisions from low to 
Fermi energies.

\subsubsection*{Dissipation Dynamics}

At energies just above the Coulomb Barrier the Prompt Dynamical 
Dipole seems to be  
rather promising. It represents a fast bremsstrahlung 
radiation during the 
charge equilibration dynamics in dissipative collisions initiated 
in charge asymmetric 
entrance channels ($M.Di~Toro$). The symmetry energy acts as a 
restoring force and
the fast neutron emission, even related to the symmetry potentials, 
is ruling the
damping. 

Moreover such fast photon emission in charge-asymmetric entrance 
channels 
could also be an interesting cooling mechanism in passing from a 
warm" to a "cold"
fusion, with a related larger survival probability of the residue. This 
could be tested
with measurements of fission times (see $D.Boilley$ talk) vs entrance 
channel asymmetry.
 
It is also possible to show that the symmetry term is influencing 
the dissipative
 dynamics in deep inelastic and fusion reactions with very neutron 
rich systems.
 $Y.Iwata$ has presented some evidences of a fusion hindrance due to 
symmetry effects
for the $132Sn+132Sn$ reaction compared to less exotic cases. 

\subsubsection*{Fast Nucleon Emission}

The isospin content of the fast nucleon emission should reveal a 
good sensitivity
to the symmetry potentials. $B.Tsang$ has show some recent data from 
MSU in reactions
$(124,124)$ and $(112,112)$ Sn on Sn at $50AMeV$. In order to eliminate 
secondary decay effects
a double ratio, n-rich vs n-poor N/Z results, is analysed. Some effect 
is observed 
in the variation of the isospin signal with the nucleon kinetic 
energy, but
for such studies a mass selection is  essential in order to separate 
different clusters.
A similar analysis, but for fragments in multifragmentation reactions,
has been proposed 
by $M.Di~Toro$. We have now an isospin distillation effect, typical 
of a two component
liquid-gas phase transition, coupled to a radial flow. The consequence 
is a symmetry
energy dependent velocity distribution of the produced fragments.

\subsubsection*{Isospin Transport}

The isospin transport during the collision should be very sensitive 
to the symmetry energy
at the density of the medium (diffusion term) and to its density 
gradient (drift term).
Therefore several related measurements have been proposed. The direct 
Isospin Equilibration
dynamics at the Fermi energies has been addressed by 
$B.Tsang$ $M.Di~Toro$ and $L.W.Chen$
through the study of the "Imbalance Ratio" at the $PLF/TLF$ rapidity. 
This is an evaluation 
of the isospin content of the isospin asymmetric collision normalized to 
the corresponding
symmetric cases \cite{rami00}. In this way the secondary decay 
effects are expected to be much reduced.
The centrality dependence of such quantity appears sensitive to the 
stiffness of the symmetry
energy, of course self-consistent transport calculations are needed. A way 
to direct 
measure this effect could be to analyse the correlation to the
amount of kinetic energy dissipation. 

$L.W.Chen$ has focussed the 
attention
on the uncertainties coming from the momentum dependence of the 
effective interactions,
that are reducing the interaction times, and from the in medium 
corrections to the nucleon
nucleon cross sections. Anyway some first conclusions seem to 
converge on a rather
stiff behavior of the symmetry term around saturation, with a 
density dependence
$E_{sym} \approx (\rho/\rho_0)^\gamma$ with $0.7 < \gamma < 1.1$.
Since the asymmetric matter compressibility is also affceted by such 
slope, \cite{baranPR}, 
an interesting cross-check comes from very recent monopole data
in the Sn isotopic chain \cite{garg07}.

The neutron enrichement
of mid-rapidity fragments for ternary events (neck fragmentation) 
seems to confirm
this relatively large gradient of the symmetry term around saturation. 
This point is not 
trivial since
a large class of effective interactions can be dropped and more 
diffuse neutron skin 
distributions in neutron excess systems can be expected.

\subsubsection*{Neutron Skin}

Unfortunately the neutron skin thickness cannot be directly tested 
by elementary probes.
$Z.X.Li$ has suggested a new observable, the proton-nucleus reaction 
cross section at
intermediate energies, that should be rather sensitive to the neutron 
distribution. The 
argument is based on Isospin-QMD transport simulations,
 but the physics behind appears very clear. As already 
noted a stiff symmetry
energy implies a larger n-thickness but together with a less 
attractive proton potential.
The result will be a smaller reaction cross section for proton nucleus 
collisions when
the N/Z of the target is increasing. Systematic measurements for 
isotopic chains
are needed.

$D.Q.Fang$ is proposing the use of the Statistical 
Abrasion Ablation model,
 limited to PLF neutron abrasion events, to reconstruct the neutron 
skin structure of
the target. 

\subsubsection*{Collective Radial Flow}

$F.S.Zhang$ has tried to see a symmetry energy dependence of the threshold 
of collective
flows, in particular the radial flow, in collisions at the Fermi energies. 
The threshold is
related to the onset of compression-expansion dynamics, which is ruled by 
the stopping power
of the NN cross sections and the nuclear matter compressibility. The latter 
is depending on 
the symmetry energy (slope and curvature) but for the asymmetries tested with 
stable isotopes
the effect appears neglgible (of the order of $1\%$ for a large 
variation of the 
symmetry
stiffness). Much more important is the dependence on the used in-medium cross 
sections.

\subsubsection*{Isoscaling}

The use of the Isoscaling signal in order to extract the symmetry energy at 
low densities
has been presented by $W.D.Tian$, $M.Veselsky$ and $W.Trautmann$. It consists 
in the
relation $R_{21}=Y_2(N.Z)/Y_1(N,Z)=C exp(\alpha N+\beta Z)$ where $Y_i(N,Z)$ 
are the yields
for the $(N,Z)$ fragment measured in two different reactions, n-rich vs 
n-poor. 

The connection
to the symmetry energy is clear in an equilibrium gran canonical
model of the multifragmentation. The relation is not so evident in a 
dynamical
approach where the isoscaling parameters appear also depending on the 
widths of
the primary isotopic and isotonic distributions \cite{liutx04}. The 
fact that the isoscaling signal
has been observed in a very different fragment production mechanisms,
 from 
deep-inelastic to fission, as well as in events where fragments are 
showing
clear dynamical features (like neck fragmentation) seems to suggest some 
caution in the 
direct use of statistical approaches.  

Some new projectile fragmentation 
data have been
shown by $W.Trautmann$ where the isospin asymmetry is enhanced by the use 
of unstable
beams (107Sn and 124La) vs the n-rich reference system 124Sn. A clear 
centrality 
dependence of the isoscaling parameters
is observed: the $\alpha$ is decreasing for more central events indicating 
less n-rich
fragments. This can be due to a neutron distillation during the primary 
fragment 
formation as well as to larger neutron sequential emission. Neutron detection 
as well
as a reconstruction of the primary fragments seem to be very important.

Finally the effect of isospin momentum dependent forces on the neutron 
distillation
has been discussed by $J.Xu$. We see a slight increase of the mixed phase 
region
where the distillation takes place. The poorly known temperature 
dependence of the 
symmetry energy could explain this result.

\subsection*{Probing the Symmetry Energy at Supra-Saturation}

The way to test the behavior of the Symmetry Term at high densities, 
of large
importance for Neutron Star Modeling, has been addressed by 
$L.W.Chen$ and $M.Di~Toro$.
Here we need beam energies at least of few hundred AMeV in 
order to reach
regions of 2-3 times $\rho_0$ during the compresssion stage. Therefore 
transport 
simulations should have at least a relativistic kinematics. In fact we 
can see that
even a fully covariant treatment of the interactions is very relevant 
 \cite{baranPR,greco03,ferini06}.
 
There is a general agreement on the observables more sensitive to the 
stiffness
of the symmetry energy in this density range, but very few data are 
presently available.
A general comment is that we need high transverse momentum selections 
when we try
to see reaction products coming from the transient high density stage. 
This largely reduces the
multiplicities... and the happiness of the experimentalists. 

Good obervables appear to be
neutron/proton (or triton/3He) ratios and flow differences. For the 
latter we expect
even an important fully relativistic effect due to the Lorentz forces, 
present in the
relativistic dynamics, that enhance the contribution of the 
vector-isovector part
of the interaction \cite{greco03}.

Rather promising seems to be the study of yield ratios of mesons 
with different
isospin structure, like $\pi^-/\pi^+$ and $K^0/K^+$. In fact the isospin 
effects on the
inelastic vertices are directly related to the nucleon/meson 
self-energies and so
the corresponding rates will be very sensitive to the isovector 
covariant 
structure
of the symmetry term. This is particularly true around the 
production thresholds
\cite{ferini06}. But we cannot ask a $p_t$ selection over there......

Certainly highly isospin asymmetric unstable beams in this energy range 
will largely
improve the overall experimental sensitivity.

Increasing the beam energy we can reach higher baryon densities but 
many new 
inelastic channels
are opening, with increasing uncertainties on the relative cross sections. 
An interesting contribution in the direction of the possibility of 
observing
new phenomena has been delivered by $B.S.Zuo$. Focussing on data from 
$pp$ collisions
he has shown the importance for the strangeness production of the 
components
beyond the 3-quark structure of the nucleon resonances, in particular 
for $N^*(1535)$ and $\Delta^*(1620)$. It was an hadron structure talk
but we can easily see the implications for heavy ion collisions at 
intermediate energies. 
 
\subsection*{Fragmentation Simulations for Cancer Therapy}
At the end of this section mainly devoted to fragmentation results 
I would like to mention
the contribution of $B.A.Bian$ with some results of fragment 
production simulations
for 12C and 20Ne beams at Fermi and Intermediate enegies on targets 
of interest
for HIC cancer therapy. This kind of studies are extremely important
 and the topic 
should require
much more attention in the near future, even considering the 
improvements of the
transport approaches. We need also more data to compare with. Moreover 
the angular distributions
 of the reaction products will be very important. E.g. it would be nice 
to check if
 prompt break-up events at forward angles can be correctly accounted for 
within
 the mean field plus fluctuation model discussed here.

\section{Particle correlations and Collective Flows}

\subsubsection*{Imaging}

As already noted before, the reconstruction of primary sources 
of the final
reaction products is of fundamental importance. $G.Verde$ has 
shown the high
degree of accuracy reached with the measurements of 
particle-particle correlation.
Of course the "Imaging" technique is always based on the 
inversion of an 
integral equation and so experimental uncertainties are always present. 
Some
very nice new results have been presented: i) on effects of the symmetry 
energy stiffness on
 pp correlations 
(in a $soft$ symmetry term the protons are more attracted); 
 ii) on a direct study of emission hierarchy, (different 
particles and clusters
 probe different sources, as expected by dynamical models); 
iii) on the 
competition with temperature effects. 

The interest even for 
spectroscopic studies
has been shown from the analysis of 12C projectile 
fragmentation data. 
$\alpha$-correlations in excited states point to resonance 
structures, in particular
to the famous Hoyle Resonance, of fundamental importance for the 
synthesis of
the elements and the formation of organic matter.

The short emitting times detected from pion interferometry 
in Au+Au collisions at the
ultrarelativistic RHIC energies have inspired the Granular 
QGP model presented
by $W,N.Zhang$. The saturation of the elliptic flows at 
high transverse momentum
is also accounted for in the same picture, although a decrease 
for larger $p_t$ is
predicted, not observed so far. The cluster structure of the 
expanding QGP
is suggested even in hadronization models via a spinodal mechanism 
in a first order
phase transition scheme (see the $M.Velsesky$ report).

\subsubsection*{Scaling of Collective Flows}

A thorough analysis of anisotropic collective flows has been 
delivered by $Y.G.Ma$,
 ranging from Fermi to ultrarelativistic energies. The stimulating 
suggestion is that
similar scaling properties should be observed in the two very different 
energy regimes
just comparing hadron flows vs. quark composition with respect to light 
ion flows
vs.  nucleon composition. The similarity is based on a common 
coalescence mechanism
for the hadron/ion formation. The idea is very suggestive. 

In fact at RHIC energies
such scaling has been observed for hadron flows in Au+Au 
collisions at 200GeV/c.
At the Fermi energies no evidence is presently available. 
$Y.G.Ma$ suggests to revisite
the available flow data accumulated in the last 20 years. 
His point is based on 
light ion ($A\leq 4)$ results derived from $BUU$
transport simulations for 86Kr+124Sn collisions at $25AMeV$, 
stopped at 200fm/c,
 very early for this energy region. Unfortunately the low energy 
heavy ion dynamics
 is much more complicated and this interesting signal can be easily 
washed out
 by later cluster emissions of different nature, dynamical as well as 
statistical.
 In any case the idea is stimulating, maybe for cluster formation at 
higher energies,
 where the final state interactions can be less important.
 A second message of this interesting talk is too look at the 
hard gamma flows, always
 in the Fermi energy domain. An anticoincidence with 
neutron/proton flows is expected
 for the high energy part of the gamma spectrum, of non thermal origin.

\section{Looking for SHE}

I feel a profound admiration for the very brave people trying 
to access a fusion dynamics
at the level of the pico-barn cross sections. We have seen the problems and 
the status of the art
in the nice talks of $Y.Abe$, $E.G.Zhao$ and $D.Boilley$. All the 
approaches are 
based
on a Brown-Motion inspired treatment, with large 
dissipation-fluctuation effects. 
In this respect it has been rather impressive the picture of the 
fusion-fission trajectories
in the deformation-separation plane shown by $Abe-san$. Since we are 
facing a two-steps
transport problem, coulomb and shape barriers, it is clear that the 
fluctuation dynamics can give
a substantial contribution. The main problem is the choice of the 
relevant degrees of freedom.

We need reactions with large neutron excess to better enter the 
expected stability
island (Z between 114 and 126, N between 172 and 184, depending of models). 
In this respect
the use of unstable neutron rich elements could be helpful, but the effect
of neutron abundance on the residue cross section is not yet clear, 
see the $E.G.Zhao$
report.   

In my opinion a very good step forward in the direction of a 
better testing of various 
fusion-fission models is related to the possibility of measuring 
long fission times using the Blocking Technique in Crystals, as 
shown by $D.Boilley$
\cite{morj07}. The fact that it has been possible to discriminate 
between 238U+Ni (Z=120 compound)
and 238U+Ge (Z=124 compound) reactiions, with evidence of long fission 
time events 
($\tau > 10^{-18}s$),
 at variance with the 208Pb+Ge (Z=114 comp.) case, represents a very 
promising result. Of course
there are limitations for the target choice, in particular the N/Z selection.

\section{Short Notes}

I have appreciated same "special" Characters during the Conference:
\begin{itemize}
\item{
$Y.M.Zhao$: {\em The Highly Random}

He suggests a nuclear spectroscopy based on Random Interactions}
\item{
$T.Maruyama$: {\em The Highly Mixed}

He sees Mixed Phases everywhere, with very exotic shapes}
\item{
$G.Verde$: {\em The Highly Correlated}

He can disentangle correlation effects in any possible system}
\item{
$Y.G.Ma$: {\em The Highly Scaled}

Scaling is always there, in any response of a complex system}
\end{itemize}

\section{Conclusions}

I hope that I have been able to express my deep feeling that
we have enjoyed a very lively and stimulating meeting, in spite of the fact 
that certainly I have missed several important points raised during
the presentations and the discussions. The very friendly atmosphere of
the Beijing Normal University and the warm support of the Physics Department
Group, lead by Prof.F.S.Zhang, have largely contributed to the success
of the Conference.

Finally I have still to remark the very nice contributions
of several young people from many different Chinese Universities 
and Research Centers.
This is the best way to show the excellent development of the field in China.
In this respect I think important to finish with the very best wishes
for the new Lanzhou Cooling-Storage-Ring facility. We have heard from Prof.
$J.S.Wang$ that everything is going on very successfully: before the end of
June 2007 three beams have been extracted, stored, cooled and accelerated up
to 1AGeV, $12C(210^9pps)$, $36Ar(1.410^9pps)$ and $129Xe(10^8pps)$.
Congratulations and looking forward to a very exciting physics!



\end{document}